\documentclass[aps,prb,showpacs,twocolumn,a4paper]{revtex4}
\usepackage{graphicx}

\begin{document}

\title{Extended staggered-flux phases in two-dimensional lattices}
\author{Yi-Fei Wang$^1$ Chang-De Gong$^{2,3,4}$ and Shi-Yao Zhu$^3$}
\affiliation{$^1$National Laboratory of Solid State
Microstructures and Department of Physics,
Nanjing University, Nanjing 210093, China\\
$^2$Chinese Center of Advanced Science and Technology (World
Laboratory),
P.O. Box 8730, Beijing 100080, China\\
$^3$Department of Physics, Hong Kong Baptist University, Kowloon
Tong, Hong Kong, China\\
$^4$Department of Physics, The Chinese University of Hong Kong,
Hong Kong, China}
\date{\today}

\begin{abstract}
Based on the so called $t$-$\phi$ model in two-dimensional (2D)
lattices, we investigate the stabilities of a class of extended
staggered-flux (SF) phases (which are the extensions of the
$\sqrt{2}\times\sqrt{2}$ SF phase to generalized spatial periods)
against the Fermi-liquid phase. Surprisingly, when away from the
nesting electron filling, some extended-SF phases take over the
dominant SF phase (the $\sqrt{2}\times\sqrt{2}$ SF phase for the
square lattice, a $1\times\sqrt{3}$ SF phase for the triangular
one), compete with the Fermi-liquid phase in nontrivial patterns,
and still occupy significant space in the phase diagram through
the advantage in the total electronic kinetic energies. The
results can be termed as the generalized Perierls
orbital-antiferromagnetic instabilities of the Fermi-liquid phase
in 2D lattice-electron models.
\end{abstract}

 \maketitle

\section{Introduction}

An intriguing orbital-current-carrying state in the square
lattice, the staggered-flux (SF) phase\cite{Affleck} (also known
as orbital antiferromagnet\cite{Halperin}, $d$-density
wave\cite{Chakravarty} state, or charge flux phase) which has the
spatial period $\sqrt{2}\times\sqrt{2}$, has been the focus of
consideration for a long time. This $\sqrt{2}\times\sqrt{2}$ SF
phase was first considered in excitonic insulators\cite{Halperin},
then in high-$T_c$ superconductors\cite{Affleck}. The absence of
experimental vindications made it discarded in favor of other
conventional orders. Recently, it regains new attentions since
unusual experimental findings in two systems: a hidden order in
the heavy-fermion compound URu$_{2}$Si$_{2}$\cite{Chandra}; a
pseudogap phenomena (See the review by Timusk {\it et
al.}\cite{Timusk}) in the underdoped region of high-$T_c$
cuprates\cite{Chakravarty}.

In parallel with the experimental contexts, numerical signatures
in the $t$-$J$ model\cite{Ivanov,Leung} and evidences in two-leg
ladder models (hall-filled $t$-$U$-$V$-$J$\cite{Marston}, doped
$t$-$J$-$V$-$V'$ and
$t$-$J_{\bot}$-$U$-$V_{\bot}$\cite{Schollwock}) of its existence
have been found. It is also being discussed now in many new
theoretical contexts: the double-exchange model\cite{Yamanaka} of
colossal-magnetoresistance manganites; the repulsive SU($N$)
Hubbard Model\cite{Honerkamp} of ultracold fermionic atoms in
optical lattices; models with ring-exchange
interactions\cite{Chung}.

For the triangular lattice case, a $1\times\sqrt{3}$ SF phase with
$\pi/2$-flux per plaquette has been considered as the mean-field
ansatz\cite{TKLee} for the Heisenberg model. Recently found
layered sodium cobalt oxide system
Na$_x$CoO$_2$$y$H$_2$O\cite{Takada} makes it a reasonable
reference state for a range of electron filling\cite{Baskaran}.

To the best of our knowledge, the formal exploration of other
possible staggered-flux phases has never been made in previous
studies (except a suggestion in studying the double-exchange
model\cite{Yamanaka}). We would address the following problems:
can a class of extended-SF phases with generalized spatial
periods, which are still macroscopically space-inversion and
time-reversal invariant, be stabilized in a 2D lattice-electron
model? if so, how do these extended-SF phases compete with each
other and with the Fermi-liquid phase? is this kind of
stabilization a generic feature of 2D lattice-electron models?

In this report, we try to answer the above questions through
numerical studies of a class of extended-SF phases in the
so-called $t$-$\phi$ model\cite{Harris} which is the simplest one
being able to generate the $\sqrt{2}\times\sqrt{2}$ SF phase
spontaneously. This model has also been shown\cite{Morse} to be
closely related to the large-$N$ SU($N$) $t$-$J$ model (where the
$\sqrt{2}\times\sqrt{2}$ SF phase is first realized as a ground
state\cite{Affleck}) and the weak-coupling Hubbard model. If the
idea about extended-SF phases can be realized in this simple
model, then it should be further taken into account and be tested
in the future studies of other more realistic models.

For each extended-SF phase, we first investigate the variation of
the total kinetic energy (TKE) of tight-binding electrons
(electron-hopping part of the $t$-$\phi$ model) versus the flux
parameter $\phi$. Further, including the ``magnetic'' energy of
the flux itself, quantum phase transitions among these extended-SF
phases and the Fermi-liquid phase are discussed in details through
quantum phase diagrams.

\section{Notation and Formulation}

\begin{figure}[!htb]
  \centering
  \vspace{0in}
  \hspace{0.1in}
  \includegraphics[scale=0.4]{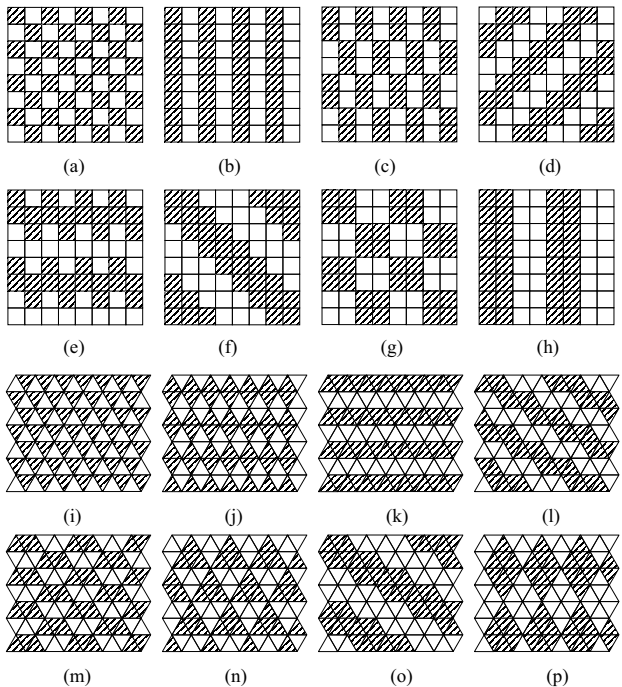}
  \vspace{-1.1in}
  \caption{Illustration of extended-SF phases in 2D lattices. Shaded plaquette has a flux $\phi$,
  and the blank one has a flux $-\phi$.
  (a)$\text{SF}^{\text{S}}_{\sqrt{2}\times\sqrt{2}}$;
  (b)$\text{SF}^{\text{S}}_{1\times2}$;
  (c)$\text{SF}^{\text{S}}_{2\times\sqrt{5}}$;
  (d)$\text{SF}^{\text{S}}_{\sqrt{2}\times2\sqrt{2}}$;
  (e)$\text{SF}^{\text{S}}_{2\times4}$;
  (f)$\text{SF}^{\text{S}}_{\sqrt{2}\times3\sqrt{2}}$;
  (g)$\text{SF}^{\text{S}}_{2\sqrt{2}\times2\sqrt{2}}$;
  (h)$\text{SF}^{\text{S}}_{1\times4}$;
  (i)$\text{SF}^{\text{T}}_{1\times1}$;
  (j)$\text{SF}^{\text{T}}_{1\times\sqrt{3},\alpha}$;
  (k)$\text{SF}^{\text{T}}_{1\times\sqrt{3},\beta}$;
  (l)$\text{SF}^{\text{T}}_{\sqrt{3}\times2}$;
  (m)$\text{SF}^{\text{T}}_{\sqrt{3}\times\sqrt{3}}$;
  (n)$\text{SF}^{\text{T}}_{2\times2}$;
  (o)$\text{SF}^{\text{T}}_{\sqrt{3}\times3}$;
  (p)$\text{SF}^{\text{T}}_{2\times2\sqrt{3}}$.}
 \label{f.1}
\end{figure}

The notation for an extended-SF phase is generally written as
$\text{SF}^{\text{L}}_{r\times s}$, where $\text{L=S,T}$ represent
the square lattice and the triangular one, respectively, and
$r\times s$ represents the spatial period ({\it i.e.}, the size of
a selected unit cell). Sometimes a Greek letter is added in the
notation to distinguish among the extended-SF phases which have
the same spatial period({\it e.g.}
$\text{SF}^{\text{T}}_{1\times\sqrt{3},\alpha}$ and
$\text{SF}^{\text{T}}_{1\times\sqrt{3},\beta}$). We will consider
$8$ kinds of extended-SF phases in each lattice. The flux
configurations of various extended-SF phases are illustrated in
Fig.~\ref{f.1}. These extended-SF phases are chosen by examining
their symmetries and simplicities. More complicated extended-SF
phases can also be obtained by the generalized routines, however,
they have less importance both theoretically and experimentally.

The following U(1) gauge-invariant tight-binding Hamiltonian
describes the $t$-$\phi$ model\cite{Harris}:
\begin{equation}
H=-t\sum_{\langle{ij}\rangle\sigma}\left(e^{i\phi_{ij}}c^{\dagger}_{j\sigma}c_{i\sigma}+\text{H.c.}\right)+{1\over2}K\sum_{p}\phi^2_p.
\label{e.1}
\end{equation}
Here $\phi_{ij}$ is in units of $\phi_0/2\pi$ ($\phi_0=hc/e$ is
the flux quantum). The nearest-neighbor hopping integral of
electrons $t$ is modified as $t_{ij}=t\exp(i\phi_{ij})$ due to the
Aharonov-Bohm effect. The second term of Eq.~(\ref{e.1}) gives the
``magnetic'' energy. $\phi_p$ is the sum of $\phi_{ij}$
(clockwisely) along a plaquette $p$. To capture the major physics,
the spatial fluctuation of $|\phi_p|$ is neglected.

For each extended-SF phase, after taking a specific gauge
(physical quantities are gauge-invariant as demonstrated
before\cite{JAn}), the first part of Eq.~(\ref{e.1}) can be
converted to $k$-space and then be directly diagonalized, leads to
several subbands of the energy spectrum. The symmetries of the
extended-SF phases in Fig.~\ref{f.1} ensure that the number of
subbands are not too large. Among these extended-SF phases, the
energy spectra of three simpler ones,
$\text{SF}^{\text{S}}_{\sqrt{2}\times\sqrt{2}}$\cite{Harris,JAn},
$\text{SF}^{\text{T}}_{1\times1}$ and
$\text{SF}^{\text{T}}_{1\times\sqrt{3},\alpha}$\cite{Wang}, have
been clarified in details before.

After getting the energy spectrum (numerically for large
Hamiltonian matrices), the total density of states (DOS)
$D(\omega)$ can be obtained by summing up the contributions from
all subbands. Thus the TKE per site for electron-filling factor
$\nu$ at zero temperature is given by
\begin{equation}
E_{kin}(\phi,\nu)=\int^{\mu}_{\omega_{min}}d \omega
D(\omega)\omega.\label{e.2}
\end{equation}
And the chemical potential $\mu$ is determined through
\begin{equation}
\nu=\int^{\mu}_{\omega_{min}}d \omega D(\omega).\label{e.3}
\end{equation}
Where $\omega_{min}$ is the lower limit of the energy spectrum.
Both the DOS and TKE are invariant under the transformations
$\phi\rightarrow-\phi$ and $\phi\rightarrow 2\pi-\phi$. Therefore,
it is enough to consider $E_{kin}$ for $0\leq\phi\leq\pi$. For
$\phi=0$ (the Fermi-liquid phase) or $\pi$ (also called the
$\pi$-flux phase\cite{Affleck} in the square lattice), all
extended-SF phases in a 2D lattice are equivalent and hence have
the same $E_{kin}$.

Then through the self-consistent determination of the flux
parameter $\phi$ by minimizing the total energy per site
determined by the $t$-$\phi$ model (TKE plus the ``magnetic''
energy), we can plot the phase diagram in the $\nu$-$K$ parameter
space by labelling the lowest-energy phase (an extended-SF phase
or the Fermi-liquid phase) in the $\nu$-$K$ parameter space.
Therefore for a given electron-filling factor $\nu$, quantum phase
transitions may occur when $K$ changes, and vice versa.

\section{Square lattice}

\begin{figure}[!htb]
  \vspace{-0.15in}
  \hspace{-0.42in}
  \includegraphics[scale=0.8]{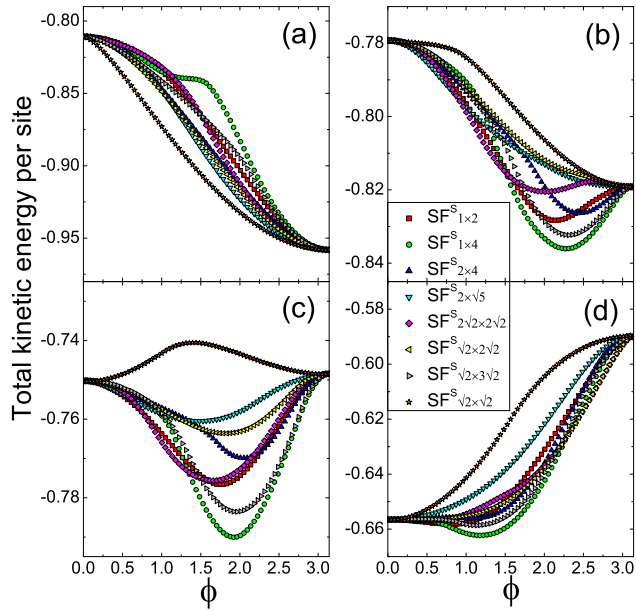}
  \vspace{-0.4in}
  \caption{(color online).  Square lattice: total kinetic energy
  per site (in units of $t$) of extended-SF phases versus
the flux parameter $\phi$ (in units of $\phi_0/2\pi$) for several
values of electron filling $\nu$'s. (a) $\nu=1/2$; (b) $\nu=3/8$;
(c) $\nu=1/3$; (d) $\nu=1/4$.} \label{f.2}
\end{figure}

In Fig.~\ref{f.2} we plot the TKE per site versus the flux
parameter $\phi$ for several values of electron filling $\nu$'s.
Due to the particle-hole symmetry in the square lattice, the DOS
always has the property $D(\omega)=D(-\omega)$, and we have
$E_{kin}(\phi,\nu)=E_{kin}(\phi,1-\nu)$. Hence $E_{kin}$-$\phi$
curves are plotted only for $\nu\leq1/2$.

At $\nu=1/2$ (Fig.~\ref{f.2}(a)),
$\text{SF}^{\text{S}}_{\sqrt{2}\times\sqrt{2}}$ has lower TKE than
the Fermi-liquid phase ($\phi=0$) and all other extended-SF phases
for a given $\phi$. That is why
$\text{SF}^{\text{S}}_{\sqrt{2}\times\sqrt{2}}$ can be stabilized
before\cite{Affleck,Harris} near half filling. While at $\nu=3/8$
(Fig.~\ref{f.2}(b)) and even smaller filling factors,
$\text{SF}^{\text{S}}_{\sqrt{2}\times\sqrt{2}}$ has higher TKE
than all other extended-SF phases for a given $\phi$. At
$\nu=3/8$, for $0<\phi<0.83$, $0.83<\phi<1.51$, and
$1.51<\phi<3.14$, $\text{SF}^{\text{S}}_{2\times\sqrt{5}}$,
$\text{SF}^{\text{S}}_{2\sqrt{2}\times2\sqrt{2}}$ and
$\text{SF}^{\text{S}}_{1\times4}$ have the lowest TKE,
respectively.

The nesting property of the Fermi surface in the Fermi-liquid
phase at $\nu=1/2$ makes the system resemble a one-dimensional
Perierls system in which a lattice distortion lowers the TKE of
electrons by opening a gap near the Fermi energy. Therefore the
advantage of extended-SF phases versus the Fermi-liquid phase in
TKE can be considered as the generalized Perierls
orbital-antiferromagnetic instabilities\cite{Harris} of the
Fermi-liquid phase near the nesting filling, similar to the
instability of lattice electrons in a uniform magnetic
field\cite{Hasegawa}. While away from the nesting filling, the
advantage of extended-SF phases in TKE is weakened gradually. As
for $\nu<1/4$, nearly all extended-SF phases have higher TKE than
the Fermi-liquid phase.

\begin{figure}[!htb]
  \vspace{-0.2in}
  \hspace{0.1in}
  \includegraphics[scale=0.7]{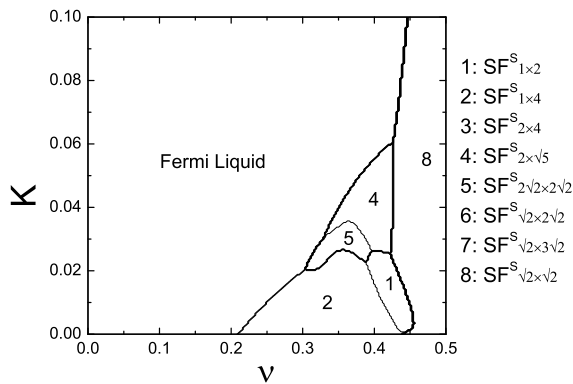}
  \vspace{-0.2in}
  \caption{Square lattice: phase
diagram in the $\nu$-$K$ parameter space ($K$ is in units of
$t$).}
  \label{f.3}
\end{figure}

The phase diagram in the $\nu$-$K$ parameter space is shown in
Fig.~\ref{f.3}. We notice that $\text{SF}^{\text{S}}_{2\times4}$,
$\text{SF}^{\text{S}}_{\sqrt{2}\times2\sqrt{2}}$ and
$\text{SF}^{\text{S}}_{\sqrt{2}\times3\sqrt{2}}$ occupy no space
in the phase diagram. At $\nu<0.21$, the Fermi-liquid phase is
robust against all extended-SF phases considered. When $\nu$
increases, $\text{SF}^{\text{S}}_{1\times4}$ appears in the
$K<0.03t$ region. For $0.3<\nu<0.425$ and $K<0.06t$,
$\text{SF}^{\text{S}}_{1\times2}$,
$\text{SF}^{\text{S}}_{1\times4}$,
$\text{SF}^{\text{S}}_{2\times\sqrt{5}}$ and
$\text{SF}^{\text{S}}_{2\sqrt{2}\times2\sqrt{2}}$ compete and show
several tricritical points. At $K>0.06t$, only
$\text{SF}^{\text{S}}_{\sqrt{2}\times\sqrt{2}}$ competes with the
Fermi-liquid phase near half filling. If more complicated
extended-SF phases are taken into account, the phase space will be
divided into smaller portions, however, the general features
should not be altered fundamentally.

The previous numerical
studies\cite{Ivanov,Leung,Marston,Schollwock} suggest that the
$\text{SF}^{\text{S}}_{\sqrt{2}\times\sqrt{2}}$ phase is a key
ingredient of the ground state in the models of strongly
correlated electrons, at least in the low doping regime (doping
$x<10\%$, {\it i.e.}, $0.45<\nu<0.5$), is in agreement with our
results. And our results also suggest that when $\nu\leq0.425$
(doping $x\geq15\%$), the
$\text{SF}^{\text{S}}_{\sqrt{2}\times\sqrt{2}}$ phase does not
predominate; when $0.425<\nu<0.45$ (doping $10\%<x<15\%$),
stripe-like extended-SF phases, $\text{SF}^{\text{S}}_{1\times2}$
and $\text{SF}^{\text{S}}_{1\times4}$ should be taken as
additional competitive ingredients of the ground state. Of course,
in order to give quantitative evaluation of the TKE or the total
free energy of interacting electrons, the hopping integral $t$
should be renormalized by many-body effects.

It has been shown before that the instabilities of the
Fermi-liquid phase in the $t$-$\phi$ model at nonzero $K/t$ can be
related to those in the large-$N$ SU($N$) $t$-$J$ model at nonzero
$t/J$\cite{Morse}. Therefore the phase diagram in Fig.~\ref{f.3}
which shows different Perierls orbital-antiferromagnetic
instabilities at different electron fillings can be used to
predict the existence of similar instabilities in the latter
model.

\section{Triangular lattice}

\begin{figure}[!htb]
  \vspace{-0.25in}
  \hspace{-0.47in}
  \includegraphics[scale=0.82]{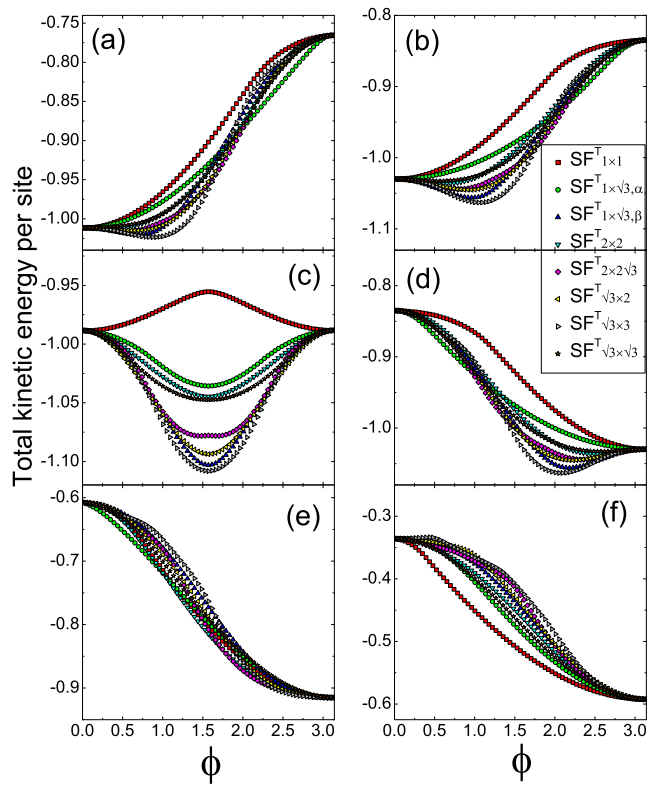}
  \vspace{-0.37in}
  \caption{(color online). Triangular lattice: total kinetic energy
  per site (in units of $t$) of extended-SF phases versus the
flux parameter $\phi$ (in units of $\phi_0/2\pi$) for various
values of electron filling $\nu$'s. (a) $\nu=1/3$; (b) $\nu=3/8$;
(c) $\nu=1/2$; (d) $\nu=5/8$; (e) $\nu=3/4$; (f) $\nu=7/8$.}
\label{f.4}
\end{figure}

For the triangular lattice, the DOS has the symmetry
$D(\omega,\pi-\phi)=D(-\omega,\phi)$\cite{Wang}, consequently,
$E_{kin}(\phi,\nu)=E_{kin}(\pi-\phi,1-\nu)$. Around the nesting
filling $\nu=3/4$, the extended-SF phases also exhibit the generic
advantage in TKE versus the Fermi-liquid phase in broad
$\nu$-$\phi$ parameter space (see Fig.~\ref{f.4}).

\begin{figure}[!htb]
  \vspace{-0.5in}
  \hspace{-0.42in}
  \includegraphics[scale=0.8]{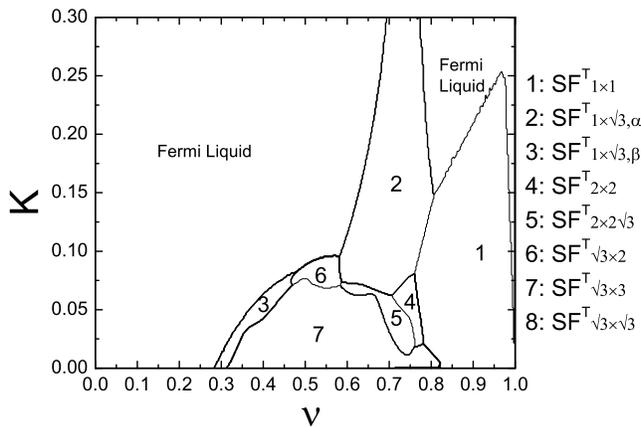}
  \vspace{-0.4in}
  \caption{Triangular lattice: phase diagram in
the $\nu$-$K$ parameter space ($K$ is in units of $t$). }
  \label{f.5}
\end{figure}

To obtain the phase diagram as shown in Fig.~\ref{f.5}, one should
notice that the ``magnetic'' energy per site is $K\phi^2$ in the
triangular lattice (in which the number of plaquettes doubles the
number of sites). $\text{SF}^{\text{T}}_{\sqrt{3}\times\sqrt{3}}$
does not appear in the phase diagram. For $\nu>0.82$ and
$K<0.26t$, $\text{SF}^{\text{T}}_{1\times1}$ dominates the phase
space. For $K<0.08t$ and $\nu$ close to the nesting filling
$\nu=3/4$, $\text{SF}^{\text{T}}_{2\times2}$,
$\text{SF}^{\text{T}}_{2\times2\sqrt{3}}$ and
$\text{SF}^{\text{T}}_{\sqrt{3}\times3}$ compete. When $\nu$ is
far away from the nesting filling ($\nu<0.27$), the Fermi-liquid
phase is robust against all extended-SF phases considered. For
larger $K$ ($K>0.26t$), only
$\text{SF}^{\text{T}}_{1\times\sqrt{3},\alpha}$ is still robust
against Fermi-liquid around $\nu=3/4$. It suggests that in
parallel with the importance of
$\text{SF}^{\text{S}}_{\sqrt{2}\times\sqrt{2}}$ near half-filling
in the square lattice,
$\text{SF}^{\text{T}}_{1\times\sqrt{3},\alpha}$ should be a key
ingredient of the ground state in some lattice-electron
models\cite{Baskaran} in the triangular lattice near $\nu=3/4$.

\section{Summary and Conclusions}

In numerous works, the
$\text{SF}^{\text{S}}_{\sqrt{2}\times\sqrt{2}}$ phase competes
with other better-known phases, such as the phases of
charge-density waves, spin-density waves, superconductivity,
stripes and Fermi liquid in the square lattice. The possibilities
of other extended-SF phases in both the square lattice and the
triangular one, and the competition among these extended-SF phases
and the Fermi-liquid phase, which have never been considered
before, are the main contributions of this report.

In conclusion, there are several points to be addressed: (a) the
generic and strong electron-filling-dependent stabilization of
extended-SF phases in some parameter space of the simple
$t$-$\phi$ model is an illustration of generalized Perierls
orbital-antiferromagnetic instabilities of the Fermi-liquid phase
in 2D lattices; (b) in the square lattice, the
$\text{SF}^{\text{S}}_{\sqrt{2}\times\sqrt{2}}$ phase is
dominantly robust close to half filling, which agrees with
previous studies; (c) away from half filling, other extended-SF
phases are possible as a result of advantage in TKE against the
Fermi-liquid phase; (d) in the triangular lattice, we find the
$\text{SF}^{\text{T}}_{1\times\sqrt{3},\alpha}$ phase is robust
around the nesting filling $\nu=3/4$ (in a manner similar to the
$\text{SF}^{\text{S}}_{\sqrt{2}\times\sqrt{2}}$ phase in the
square lattice), and other extended-SF phases are also possible
when $\nu$ is away from $3/4$; (e) since lowering TKE of electrons
plays an decisive role in minimizing the total mean-field free
energy in many cases, we keep cautious but optimistic hopes about
applying our nontrivial new results based on a simple model to
other more realistic systems such as the $t$-$J$ model and its
large-$N$ limit, the Hubbard or extended Hubbard model and their
weak-coupling limits or large-$N$ limits, the double-exchange
model and models with ring-exchange interactions.

This work is supported by the Chinese National Natural Science
Foundation and RGC of HK Government, and FRG of HKBU.

\end{document}